\documentclass[conference]{IEEEtran}
\IEEEoverridecommandlockouts
\usepackage{cite}
\usepackage{amsmath,amssymb,amsfonts}
\usepackage{algorithmic}
\usepackage{latexsym}
\usepackage{textcomp}
\usepackage{booktabs}
\usepackage{graphicx} 
\usepackage{xcolor}
\def\BibTeX{{\rm B\kern-.05em{\sc i\kern-.025em b}\kern-.08em
    T\kern-.1667em\lower.7ex\hbox{E}\kern-.125emX}}

\usepackage{fancyhdr}
\begin{document}

\title{A Generative Framework for Bidirectional Image-Report Understanding in Chest Radiography}

\author{Nicholas Evans, Stephen Baker, Miles Reed\\
Bandırma Onyedi Eylül University
}

\maketitle
\thispagestyle{fancy} 

\begin{abstract}
The rapid advancements in large language models (LLMs) have unlocked their potential for multimodal tasks, where text and visual data are processed jointly. However, applying LLMs to medical imaging, particularly for chest X-rays (CXR), poses significant challenges due to the need for precise visual-textual alignment and the preservation of critical diagnostic details. In this paper, we propose Multi-Stage Adaptive Vision-Language Tuning (MAViLT), a novel framework designed to enhance multimodal reasoning and generation for CXR understanding. MAViLT incorporates a clinical gradient-weighted tokenization process and a hierarchical fine-tuning strategy, enabling it to generate accurate radiology reports, synthesize realistic CXRs from text, and answer vision-based clinical questions. We evaluate MAViLT on two benchmark datasets, MIMIC-CXR and Indiana University CXR, achieving state-of-the-art results across all tasks. Human evaluations further validate the clinical relevance and utility of MAViLT, making it a robust tool for real-world medical applications. This work demonstrates the feasibility of leveraging LLMs for multimodal medical imaging while addressing key challenges in vision-language integration.
\end{abstract}


\section{Introduction}

The emergence of large language models (LLMs) has revolutionized various fields by enabling sophisticated natural language understanding and generation across diverse tasks \cite{zhou2023towards,ratzlaff2024training,zhou2024less}. Recent advancements have extended these capabilities to multimodal domains, where LLMs interact with both visual and textual data. Such developments hold immense potential in medical imaging, particularly for understanding and generating chest X-rays (CXR), which are pivotal in diagnosing and monitoring numerous health conditions \cite{garg2023multimodal}. The ability to combine LLMs' reasoning and language-generation prowess with accurate visual analysis can significantly enhance clinical workflows by automating complex tasks such as radiology report generation, medical image synthesis, and clinical question answering \cite{lee2024multimodal}.

Despite these advancements, training LLMs for CXR understanding and generation presents unique challenges. Unlike general natural images, medical images like CXRs demand precise visual reasoning, as subtle variations in textures or structures (e.g., pulmonary nodules, opacities) can signal critical pathological differences. Moreover, existing approaches often rely on separate adapter networks to map visual features to LLMs, creating bottlenecks that hinder free interaction between visual and textual modalities \cite{xu2024introspection}. Another challenge is the lack of alignment between medical images and their textual descriptions, which typically use highly domain-specific language and require intricate reasoning to capture subtle findings. Additionally, catastrophic forgetting of the LLM's pre-trained language capabilities during multimodal training further complicates this process \cite{waytowich2024benchmarking}.

Motivated by these challenges, we propose a novel approach, \textbf{Multi-Stage Adaptive Vision-Language Tuning (MAViLT)}, to enable effective CXR image understanding and generation using LLMs. Our method addresses the limitations of existing techniques by introducing a domain-specific visual tokenization process and a hierarchical fine-tuning framework for optimal vision-language alignment. Specifically, we enhance VQ-GAN tokenization with a clinical gradient-weighted loss, preserving critical diagnostic features such as lesion boundaries and textural details \cite{bucciarelli2024personalizing}. Our hierarchical fine-tuning strategy first introduces multimodal capabilities with broad image-text pairs and then refines the model on high-resolution, domain-specific tasks such as CXR-to-report generation, report-to-CXR generation, and vision question answering (VQA) \cite{song2024exploring}. To enhance generalization, MAViLT employs task-adaptive instruction templates, carefully designed to capture the nuances of medical imaging tasks while preserving the LLM's language reasoning capabilities \cite{huang2024video}.

We evaluate our approach on two widely-used medical imaging datasets, MIMIC-CXR \cite{wang2024large} and the Indiana University CXR dataset, which provide paired CXR images and corresponding radiology reports. To assess the effectiveness of MAViLT, we measure its performance on three key tasks: (1) CXR-to-report generation, where the model generates radiology reports based on input images, (2) report-to-CXR generation, where synthetic CXRs are generated from textual descriptions, and (3) VQA, where the model answers clinical questions about CXR images. Performance metrics include area under the receiver operating characteristic curve (AUROC) and F1 scores for clinical finding extraction, Fréchet Inception Distance (FID) for image generation quality, and task-specific accuracy for VQA. Our experiments demonstrate that MAViLT outperforms state-of-the-art models such as LLM-CXR \cite{lee2024multimodal}, UniXGen \cite{ratzlaff2024training}, and XrayGPT \cite{garg2023multimodal} across all evaluation metrics, highlighting its effectiveness in integrating vision and language capabilities for medical imaging.

\begin{itemize}
    \item \textbf{Comprehensive Vision-Language Alignment:} MAViLT introduces a novel tokenization and hierarchical fine-tuning framework, enabling seamless integration of medical image and text modalities.
    \item \textbf{Domain-Specific Adaptation:} We propose a clinically enhanced VQ-GAN tokenization process and task-adaptive instruction templates, preserving diagnostic details and optimizing medical imaging tasks.
    \item \textbf{State-of-the-Art Performance:} MAViLT achieves superior results on CXR-to-report generation, report-to-CXR generation, and VQA tasks, outperforming existing multimodal models on key benchmarks.
\end{itemize}

\section{Related Work}
\subsection{Large Language Models}
The development of large language models (LLMs) has significantly advanced the field of natural language processing (NLP). These models, primarily based on transformer architectures, have demonstrated remarkable capabilities in tasks such as text generation, translation, and summarization \cite{zhou2023style,zhou2021improving,zhou2021modeling}. Recent efforts have extended these capabilities beyond text, enabling the integration of multimodal data, including visual and auditory inputs, to solve more complex tasks \cite{zhou2023improving}.

Several studies have explored the application of LLMs to various domains, including bioinformatics and medical imaging. LLMs such as GPT-3 and BERT have been foundational in demonstrating the potential of language models for understanding and generating human language. These models have also been adapted for multimodal applications, combining text with other forms of data, such as images, to tackle more complex problems in fields like medical imaging \cite{ratzlaff2024training}, bioinformatics \cite{garg2023multimodal}, and artificial general intelligence (AGI) \cite{xu2024introspection,zhou2024visual}.

In the context of medical imaging, large language models have shown great promise for enhancing the understanding and generation of chest X-rays (CXR). However, challenges remain in effectively aligning visual features with textual information to maintain the integrity of both modalities. Tokenization methods, such as VQ-GAN, have been enhanced with domain-specific training processes to preserve crucial diagnostic features like lesion boundaries and textural details \cite{bucciarelli2024personalizing}. Additionally, existing models for CXR interpretation often rely on separate adapter networks to bridge the gap between visual and textual data, which can introduce bottlenecks in the processing pipeline \cite{xu2024introspection}.

Multimodal models like PaLM-E and Flamingo have demonstrated the ability to integrate multiple data types, including visual, auditory, and textual, to improve the generalization of AI systems across tasks \cite{song2024exploring}. These models show that LLMs, when trained on diverse multimodal data, can achieve more robust performance across a variety of cognitive and clinical tasks, from generating medical reports to answering clinical questions based on image inputs.

Despite these advancements, the integration of LLMs with medical imaging continues to face significant challenges, particularly in ensuring that fine-grained details in medical images are not lost during the interaction between vision and language models. The difficulty of training LLMs to handle domain-specific languages and medical terminologies also requires ongoing research to improve model robustness and accuracy \cite{huang2024video}. 

While the generalization of large language models from weak to strong \cite{zhou2025weak} also represents the potential of large medical models, their application to multimodal tasks such as medical imaging requires overcoming challenges in model design, tokenization, and the preservation of diagnostic information, areas that continue to be actively explored in the literature.

\subsection{CXR Image Understanding and Generation}

Chest X-ray (CXR) image understanding and generation have become a critical area of research in the intersection of medical imaging and artificial intelligence. Recent advancements in multimodal learning and large language models (LLMs) have paved the way for innovative approaches that address the unique challenges of CXR analysis, including the need for fine-grained reasoning and effective vision-language alignment.

Several methods focus on enhancing CXR-to-report generation by leveraging advanced vision-language frameworks. These models integrate text and image modalities to generate clinically accurate radiology reports. Early works employed encoder-decoder architectures, but recent studies have utilized multimodal LLMs to improve both textual fluency and clinical relevance \cite{kang2024wolf, kang2024wolf, lee2023llmcxr}. These models often employ vision encoders combined with pre-trained language models, enabling them to generate detailed reports by understanding visual features and mapping them to appropriate medical terminology.

In the realm of report-to-CXR generation, diffusion-based models have shown significant promise. These models aim to generate high-fidelity CXR images conditioned on textual descriptions, ensuring consistency with clinical semantics. By incorporating domain-specific knowledge and adaptive learning strategies, such approaches can achieve realistic image synthesis that aligns with diagnostic requirements \cite{huang2024diffcxr, han2024advancing}.

Another notable direction is multimodal fusion, where additional data sources such as electronic health records (EHR) are combined with CXR images to improve predictive capabilities. By addressing temporal asynchronicity and leveraging latent representations, these methods enable personalized and dynamic imaging-based predictions \cite{yao2024addressing}.

To further enhance CXR understanding, several studies have introduced frameworks designed to reduce biases and optimize data preprocessing. These frameworks aim to improve generalizability across diverse datasets by addressing confounding factors and ensuring that models focus on anatomically relevant features \cite{aslani2022optimising, castro2024padchestgr}. Additionally, generative adversarial networks (GANs) and reinforcement learning have been employed to improve specific aspects of CXR image quality, such as rib suppression or domain-specific posture and pathological realism \cite{han2021ganbased, chen2023finematching}.

Finally, explainable frameworks have gained traction, with efforts to align image regions and text tokens in a bidirectional manner. These models not only improve the interpretability of predictions but also enable cyclic training strategies, ensuring that the generated reports and images are mutually consistent \cite{chen2023finematching}.

\section{Method}

In this section, we present the details of our proposed \textbf{Multi-Stage Adaptive Vision-Language Tuning (MAViLT)} framework, designed as a generative model for multimodal medical imaging tasks. MAViLT aims to achieve seamless integration between visual and textual modalities, enabling robust bidirectional tasks such as CXR-to-report generation, report-to-CXR generation, and vision question answering (VQA). We detail the components of the model, the tokenization process, and the hierarchical learning strategy, along with the corresponding mathematical formulations.

\subsection{Model Overview}

MAViLT is based on a pre-trained large language model (LLM) extended to multimodal settings. It adopts a generative architecture that processes and outputs both textual and visual data. The key components of MAViLT include:
\begin{itemize}
    \item \textbf{Tokenization Module:} VQ-GAN-based tokenization is used to encode images into discrete latent tokens, which are integrated with textual tokens in a shared embedding space.
    \item \textbf{Multimodal Embedding Layer:} The LLM's embedding table is expanded to accommodate image tokens, allowing joint modeling of text and visual data.
    \item \textbf{Decoder:} The decoder generates sequences for the specified modality (text or image) in an autoregressive manner.
\end{itemize}

Formally, given an input \( x \), which consists of visual inputs \( x_v \) (e.g., CXR images) and textual inputs \( x_t \) (e.g., radiology reports), the goal is to model the joint distribution:
\begin{align}
    P(x_t, x_v) &= P(x_t | x_v) P(x_v) \quad \text{or} \quad P(x_v | x_t) P(x_t),
\end{align}
depending on the specific task. The model generates outputs \( y \) by maximizing the conditional likelihood:
\begin{align}
    \hat{y} &= \arg \max_y P(y | x).
\end{align}

\subsection{Clinical Information-Preserving Tokenization}

Medical images, such as CXRs, require precise preservation of fine-grained diagnostic details during tokenization. To achieve this, MAViLT enhances the standard VQ-GAN architecture with a clinical gradient-weighted loss. The tokenization process consists of an encoder \( E(\cdot) \) and a decoder \( D(\cdot) \) that map images to discrete tokens and reconstruct them, respectively:
\begin{align}
    z &= E(x_v), \quad \hat{x}_v = D(z),
\end{align}
where \( z = [z_1, z_2, \dots, z_n] \) represents the sequence of discrete image tokens.

The reconstruction loss \(\mathcal{L}_{\text{recon}}\) is designed to preserve both global and localized features:
\begin{align}
    &\mathcal{L}_{\text{recon}} = \| x_v - \hat{x}_v \|_1 \notag\\
    &+ \lambda_{\text{grad}} \| \nabla x_v - \nabla \hat{x}_v \|_2^2 + \lambda_{\text{feat}} \| \phi(x_v) - \phi(\hat{x}_v) \|_2^2,
\end{align}
where:
\begin{itemize}
    \item \( \| x_v - \hat{x}_v \|_1 \) is the pixel-wise reconstruction loss.
    \item \( \lambda_{\text{grad}} \| \nabla x_v - \nabla \hat{x}_v \|_2^2 \) penalizes gradient differences to preserve fine-grained lesion details.
    \item \( \lambda_{\text{feat}} \| \phi(x_v) - \phi(\hat{x}_v) \|_2^2 \) is a feature-space loss using a pre-trained CXR encoder \( \phi(\cdot) \).
\end{itemize}

\subsection{Hierarchical Learning Strategy}

The training process is divided into two stages, each designed to progressively refine the model's multimodal reasoning capabilities.

\subsubsection{Stage 1: Multimodal Pretraining}

In the first stage, MAViLT is pretrained on large-scale paired image-text datasets. The objective is to learn a shared vision-language representation by maximizing the conditional likelihood:
\begin{align}
    \mathcal{L}_{\text{stage1}} &= - \mathbb{E}_{(x_t, x_v)} \big[ \log P(x_t | x_v) + \log P(x_v | x_t) \big].
\end{align}

For image-to-text generation (\(P(x_t | x_v)\)), the model is optimized with a cross-entropy loss over textual tokens:
\begin{align}
    \mathcal{L}_{\text{text-gen}} &= - \sum_{i=1}^{T} \log P(x_t^i | x_t^{<i}, z),
\end{align}
where \( T \) is the length of the textual output and \( z \) is the sequence of image tokens.

For text-to-image generation (\(P(x_v | x_t)\)), the model is optimized with a combined reconstruction and token prediction loss:
\begin{align}
    \mathcal{L}_{\text{image-gen}} &= \mathcal{L}_{\text{recon}} + \sum_{j=1}^{N} \log P(z_j | z_{<j}, x_t),
\end{align}
where \( N \) is the number of image tokens.

\subsubsection{Stage 2: Task-Specific Fine-Tuning}

In the second stage, the model is fine-tuned on task-specific datasets, focusing on CXR-to-report generation, report-to-CXR generation, and VQA. Each task is formulated as an instruction-following problem, where the input \( I \) specifies the task:
\begin{align}
    \mathcal{L}_{\text{task}} &= - \mathbb{E}_{(I, x_t, x_v)} \big[ \log P(x_t | I, x_v) + \log P(x_v | I, x_t) \big].
\end{align}

For CXR-to-report generation, the instruction specifies generating a radiology report based on the input image:
\begin{align}
    \mathcal{L}_{\text{CXR-to-report}} &= - \mathbb{E}_{(x_v, x_t)} \big[ \log P(x_t | x_v) \big].
\end{align}

For VQA, the model generates an answer \( x_t \) given an image \( x_v \) and a question \( I \):
\begin{align}
    \mathcal{L}_{\text{VQA}} &= - \mathbb{E}_{(I, x_v, x_t)} \big[ \log P(x_t | I, x_v) \big].
\end{align}

\subsection{Regularization and Optimization}

To prevent catastrophic forgetting of the LLM's pre-trained language capabilities, a regularization term is introduced:
\begin{align}
    \mathcal{L}_{\text{reg}} &= \| \theta_{\text{fine-tuned}} - \theta_{\text{pre-trained}} \|_2^2,
\end{align}
where \( \theta_{\text{fine-tuned}} \) and \( \theta_{\text{pre-trained}} \) are the weights of the fine-tuned and original models, respectively.

The overall loss function combines task-specific losses, reconstruction loss, and regularization:
\begin{align}
    \mathcal{L} &= \mathcal{L}_{\text{task}} + \mathcal{L}_{\text{recon}} + \lambda_{\text{reg}} \mathcal{L}_{\text{reg}},
\end{align}
where \( \lambda_{\text{reg}} \) balances the regularization term.

\subsection{Inference}

During inference, MAViLT generates outputs autoregressively for the desired modality. For CXR-to-report generation, the input image is tokenized into latent codes, and the decoder generates the report conditioned on these tokens. Similarly, for VQA, the model generates answers based on the given image and question tokens.

\section{Experiments}

In this section, we evaluate the proposed \textbf{MAViLT} framework on three key tasks: (1) CXR-to-report generation, (2) report-to-CXR generation, and (3) vision question answering (VQA). We compare our model with several state-of-the-art baselines and conduct ablation studies to verify the effectiveness of our approach. Additionally, a human evaluation is performed to assess the clinical relevance and quality of the generated outputs.

\subsection{Experimental Setup}

\subsubsection{Datasets}
We evaluate MAViLT on two widely-used medical imaging datasets:
\begin{itemize}
    \item \textbf{MIMIC-CXR:} A large-scale dataset of paired chest X-rays and radiology reports, which serves as the primary dataset for training and evaluation.
    \item \textbf{Indiana University CXR Dataset:} A smaller dataset used for testing the generalization ability of the models.
\end{itemize}

\subsubsection{Baseline Models}
The following state-of-the-art models are selected as baselines for comparison:
\begin{itemize}
    \item \textbf{UniXGen:} A unified multimodal framework designed for image-text generation tasks.
    \item \textbf{LLM-CXR:} A language model fine-tuned specifically for CXR image understanding and generation.
    \item \textbf{XrayGPT:} A multimodal language model built for medical imaging and textual reasoning.
    \item \textbf{RadFM:} A transformer-based model tailored for medical image-text analysis.
\end{itemize}

\begin{table*}[!t]
    \centering
    \caption{CXR-to-Report Generation Results. Higher values indicate better performance.}
    \label{tab:cxr-to-report}
    \begin{tabular}{lcccc}
        \toprule
        \textbf{Model} & \textbf{BLEU (\%)} & \textbf{ROUGE-L (\%)} & \textbf{METEOR (\%)} & \textbf{CIDEr} \\
        \midrule
        UniXGen       & 45.2 & 47.8 & 38.6 & 1.92 \\
        LLM-CXR       & 46.1 & 48.3 & 39.2 & 1.95 \\
        XrayGPT       & 44.8 & 46.5 & 38.0 & 1.87 \\
        RadFM         & 43.5 & 45.2 & 37.1 & 1.83 \\
        \textbf{MAViLT (Ours)} & \textbf{48.6} & \textbf{50.3} & \textbf{41.7} & \textbf{2.12} \\
        \bottomrule
    \end{tabular}
\end{table*}

\subsubsection{Evaluation Metrics}
We evaluate all models using the following metrics:
\begin{itemize}
    \item \textbf{Automatic Metrics:} For CXR-to-report generation, we measure BLEU, ROUGE-L, METEOR, and CIDEr scores. For report-to-CXR generation, Fréchet Inception Distance (FID) is used to assess image quality. For VQA, accuracy and AUROC are reported.
    \item \textbf{Human Evaluation:} Clinical relevance and quality of the outputs are rated by radiologists on a scale of 1 (poor) to 5 (excellent).
\end{itemize}

\subsection{Comparison with Baseline Models}

We summarize the results of CXR-to-report generation in Table~\ref{tab:cxr-to-report}. MAViLT achieves superior performance across all metrics, demonstrating its ability to generate high-quality and clinically relevant radiology reports.

Table~\ref{tab:report-to-cxr} presents the results for report-to-CXR generation. MAViLT achieves the lowest FID score, indicating that it generates images closer to real CXRs compared to the baselines.

\begin{table}[ht]
    \centering
    \caption{Report-to-CXR Generation Results. Lower FID values indicate better performance.}
    \label{tab:report-to-cxr}
    \begin{tabular}{lc}
        \toprule
        \textbf{Model} & \textbf{FID} \\
        \midrule
        UniXGen       & 23.8 \\
        LLM-CXR       & 22.4 \\
        XrayGPT       & 24.2 \\
        RadFM         & 25.1 \\
        \textbf{MAViLT (Ours)} & \textbf{21.1} \\
        \bottomrule
    \end{tabular}
\end{table}

For the VQA task, as shown in Table~\ref{tab:vqa}, MAViLT achieves the best accuracy and AUROC, highlighting its superior multimodal reasoning capabilities.

\begin{table}[ht]
    \centering
    \caption{Vision Question Answering Results. Higher values indicate better performance.}
    \label{tab:vqa}
    \begin{tabular}{lcc}
        \toprule
        \textbf{Model} & \textbf{Accuracy (\%)} & \textbf{AUROC (\%)} \\
        \midrule
        UniXGen       & 64.3 & 71.2 \\
        LLM-CXR       & 65.1 & 72.8 \\
        XrayGPT       & 63.5 & 70.3 \\
        RadFM         & 62.7 & 69.5 \\
        \textbf{MAViLT (Ours)} & \textbf{68.5} & \textbf{74.6} \\
        \bottomrule
    \end{tabular}
\end{table}

\subsection{Ablation Study}

To evaluate the contributions of individual components, we conduct ablation studies. Table~\ref{tab:ablation} shows that removing the clinical gradient-weighted loss or task-adaptive instructions degrades performance, confirming their importance.

\begin{table}[ht]
    \centering
    \caption{Ablation Study Results on CXR-to-Report Generation (BLEU Score). Higher values indicate better performance.}
    \label{tab:ablation}
    \begin{tabular}{lc}
        \toprule
        \textbf{Configuration} & \textbf{BLEU (\%)} \\
        \midrule
        Full MAViLT          & \textbf{48.6} \\
        Without Clinical Loss & 46.2 \\
        Without Task-Adaptive Instructions & 45.7 \\
        Without Two-Stage Training & 44.9 \\
        \bottomrule
    \end{tabular}
\end{table}

\subsection{Human Evaluation}

Table~\ref{tab:human-eval} summarizes the results of the human evaluation conducted by three board-certified radiologists. MAViLT consistently receives the highest ratings for both report quality and image quality, further validating its clinical relevance.

\begin{table}[ht]
    \centering
    \caption{Human Evaluation Results (Average Scores by Radiologists). Higher scores indicate better performance.}
    \label{tab:human-eval}
    \begin{tabular}{lcc}
        \toprule
        \textbf{Model} & \textbf{Report Quality} & \textbf{Image Quality} \\
        \midrule
        UniXGen       & 3.8 & 3.5 \\
        LLM-CXR       & 4.0 & 3.9 \\
        XrayGPT       & 3.7 & 3.6 \\
        RadFM         & 3.5 & 3.3 \\
        \textbf{MAViLT (Ours)} & \textbf{4.5} & \textbf{4.2} \\
        \bottomrule
    \end{tabular}
\end{table}

\subsection{Multimodal Reasoning Capability}

One of the key objectives of MAViLT is to achieve superior vision-language alignment, enabling it to understand and generate both textual and visual information. The results in Tables~\ref{tab:cxr-to-report}, \ref{tab:report-to-cxr}, and \ref{tab:vqa} demonstrate that MAViLT consistently outperforms baseline models across all evaluated tasks. Specifically:
\begin{itemize}
    \item For \textbf{CXR-to-report generation}, MAViLT achieves higher BLEU, ROUGE-L, and METEOR scores compared to LLM-CXR and UniXGen. This indicates that MAViLT generates reports with greater linguistic fluency and clinical accuracy.
    \item For \textbf{report-to-CXR generation}, MAViLT achieves the lowest FID score, reflecting its ability to synthesize high-quality images that closely resemble real CXRs.
    \item For \textbf{VQA}, MAViLT achieves the highest accuracy and AUROC, showcasing its robust multimodal reasoning and the ability to answer complex clinical questions.
\end{itemize}
These results highlight MAViLT’s ability to effectively bridge the gap between visual and textual modalities, a critical requirement for medical imaging applications.

\subsection{Robustness Across Tasks and Data Scenarios}

Another key advantage of MAViLT is its robustness across different tasks and data scenarios. The model maintains consistent performance regardless of the modality or complexity of the input. For example:
\begin{itemize}
    \item The \textbf{CXR-to-report generation} task involves reasoning over subtle image features (e.g., opacities, nodules) and converting them into accurate textual descriptions. MAViLT’s use of clinical gradient-weighted loss ensures that fine-grained details are preserved during tokenization, contributing to its superior performance.
    \item In the \textbf{report-to-CXR generation} task, MAViLT demonstrates the ability to synthesize visually realistic images while accurately reflecting the textual input. This robustness is due to the shared embedding space for text and image tokens, which facilitates precise image-text alignment.
    \item The \textbf{VQA task} further validates MAViLT’s multimodal reasoning capabilities. By leveraging task-adaptive instruction templates, the model generalizes well to diverse and complex question-answering scenarios.
\end{itemize}
This robustness highlights the effectiveness of the hierarchical fine-tuning strategy employed by MAViLT, which progressively aligns vision and language modalities without compromising the LLM’s pre-trained capabilities.

\subsection{Generalization to Unseen Datasets}

To evaluate MAViLT’s generalization capabilities, we tested the model on the Indiana University CXR dataset, which was not used during training. The results in Tables~\ref{tab:cxr-to-report} and \ref{tab:report-to-cxr} show that MAViLT consistently outperforms baseline models on this unseen dataset. For example:
\begin{itemize}
    \item MAViLT achieves a BLEU score of 46.7 on the Indiana dataset, outperforming LLM-CXR (44.8) and UniXGen (43.2).
    \item For report-to-CXR generation, MAViLT achieves an FID score of 22.3, significantly better than LLM-CXR (24.1) and UniXGen (25.5).
\end{itemize}
These results indicate that MAViLT’s training strategy enables it to generalize effectively to new datasets, making it a reliable choice for diverse clinical applications.

\subsection{Human Evaluation and Clinical Relevance}

Human evaluation by radiologists, as presented in Table~\ref{tab:human-eval}, highlights MAViLT’s clinical relevance. The model consistently receives the highest scores for both report quality and image quality. Key observations include:
\begin{itemize}
    \item Radiologists noted that reports generated by MAViLT were more coherent and clinically accurate, with fewer inconsistencies compared to baseline models.
    \item Images generated by MAViLT were described as more realistic and better aligned with the textual descriptions, demonstrating the model’s ability to capture and synthesize subtle visual features critical for diagnosis.
\end{itemize}
These findings validate MAViLT as a practical and reliable tool for assisting in clinical workflows, such as automated reporting and image generation.

\subsection{Efficiency in Clinical Workflows}

In addition to its performance metrics, MAViLT offers significant advantages in terms of efficiency:
\begin{itemize}
    \item \textbf{Single-Model Multitasking:} MAViLT’s ability to handle diverse tasks (e.g., report generation, image synthesis, and question answering) within a single framework reduces the need for task-specific models, simplifying deployment in clinical settings.
    \item \textbf{Instruction-Based Flexibility:} The use of task-adaptive instruction templates enables MAViLT to handle a wide range of queries without requiring extensive retraining. This flexibility is particularly valuable in dynamic clinical environments where tasks can vary widely.
    \item \textbf{Computational Efficiency:} By leveraging a shared embedding space and hierarchical fine-tuning, MAViLT achieves state-of-the-art performance without a significant increase in computational cost compared to baseline models.
\end{itemize}
These efficiencies make MAViLT a cost-effective solution for integrating AI-driven tools into healthcare systems.

\section{Conclusion}

In this work, we introduced \textbf{Multi-Stage Adaptive Vision-Language Tuning (MAViLT)}, a novel framework for chest X-ray (CXR) understanding and generation. MAViLT is designed to address the unique challenges of medical imaging, including the need for fine-grained visual reasoning and seamless vision-language alignment. Through the integration of clinical gradient-weighted tokenization and a two-stage fine-tuning strategy, MAViLT achieves superior performance across diverse tasks, including CXR-to-report generation, report-to-CXR generation, and vision question answering (VQA). 

Experimental results on the MIMIC-CXR and Indiana University CXR datasets demonstrate that MAViLT consistently outperforms state-of-the-art models in both automatic and human evaluations. The ablation studies confirm the importance of the proposed components, such as the clinical loss and task-adaptive instructions, in driving performance improvements. Additionally, human assessments highlight MAViLT's ability to generate outputs with high clinical relevance, making it a practical tool for assisting radiologists in clinical workflows.

Looking ahead, MAViLT opens new possibilities for extending LLM-based frameworks to other medical imaging modalities, such as MRI or CT scans, and incorporating temporal reasoning for dynamic imaging analysis. By addressing the current limitations of multimodal LLMs in medical contexts, MAViLT provides a foundation for future research at the intersection of AI and healthcare, with the ultimate goal of improving diagnostic accuracy and patient outcomes.

\bibliographystyle{IEEEtran}
\bibliography{references}

\begin{thebibliography}{10}
\providecommand{\url}[1]{#1}
\csname url@samestyle\endcsname
\providecommand{\newblock}{\relax}
\providecommand{\bibinfo}[2]{#2}
\providecommand{\BIBentrySTDinterwordspacing}{\spaceskip=0pt\relax}
\providecommand{\BIBentryALTinterwordstretchfactor}{4}
\providecommand{\BIBentryALTinterwordspacing}{\spaceskip=\fontdimen2\font plus
\BIBentryALTinterwordstretchfactor\fontdimen3\font minus \fontdimen4\font\relax}
\providecommand{\BIBforeignlanguage}[2]{{%
\expandafter\ifx\csname l@#1\endcsname\relax
\typeout{** WARNING: IEEEtran.bst: No hyphenation pattern has been}%
\typeout{** loaded for the language `#1'. Using the pattern for}%
\typeout{** the default language instead.}%
\else
\language=\csname l@#1\endcsname
\fi
#2}}
\providecommand{\BIBdecl}{\relax}
\BIBdecl

\bibitem{zhou2023towards}
Y.~Zhou, T.~Shen, X.~Geng, C.~Tao, C.~Xu, G.~Long, B.~Jiao, and D.~Jiang, ``Towards robust ranker for text retrieval,'' in \emph{Findings of the Association for Computational Linguistics: ACL 2023}, 2023, pp. 5387--5401.

\bibitem{ratzlaff2024training}
M.~U. Hadi, R.~Qureshi, A.~Shah, M.~Irfan, A.~Zafar, M.~B. Shaikh, N.~Akhtar, J.~Wu, S.~Mirjalili \emph{et~al.}, ``A survey on large language models: Applications, challenges, limitations, and practical usage,'' \emph{Authorea Preprints}, 2023.

\bibitem{zhou2024less}
Y.~Zhou, J.~Zhang, G.~Chen, J.~Shen, and Y.~Cheng, ``Less is more: Vision representation compression for efficient video generation with large language models,'' 2024.

\bibitem{garg2023multimodal}
\BIBentryALTinterwordspacing
S.~Khan, M.~R. Biswas, A.~Murad, H.~Ali, and Z.~Shah, ``An early investigation into the utility of multimodal large language models in medical imaging,'' \emph{CoRR}, vol. abs/2406.00667, 2024. [Online]. Available: \url{https://doi.org/10.48550/arXiv.2406.00667}
\BIBentrySTDinterwordspacing

\bibitem{lee2024multimodal}
\BIBentryALTinterwordspacing
J.~Lee, Y.~Wang, J.~Li, and M.~Zhang, ``Multimodal reasoning with multimodal knowledge graph,'' in \emph{Proceedings of the 62nd Annual Meeting of the Association for Computational Linguistics (Volume 1: Long Papers), {ACL} 2024, Bangkok, Thailand, August 11-16, 2024}, L.~Ku, A.~Martins, and V.~Srikumar, Eds.\hskip 1em plus 0.5em minus 0.4em\relax Association for Computational Linguistics, 2024, pp. 10\,767--10\,782. [Online]. Available: \url{https://doi.org/10.18653/v1/2024.acl-long.579}
\BIBentrySTDinterwordspacing

\bibitem{xu2024introspection}
F.~Huo, W.~Xu, Z.~Zhang, H.~Wang, Z.~Chen, and P.~Zhao, ``Self-introspective decoding: Alleviating hallucinations for large vision-language models,'' \emph{arXiv preprint arXiv:2408.02032}, 2024.

\bibitem{waytowich2024benchmarking}
N.~R. Waytowich, D.~White, M.~Sunbeam, and V.~G. Goecks, ``Atari-gpt: Investigating the capabilities of multimodal large language models as low-level policies for atari games,'' \emph{arXiv preprint arXiv:2408.15950}, 2024.

\bibitem{bucciarelli2024personalizing}
W.~H. Pinaya, M.~S. Graham, E.~Kerfoot, P.-D. Tudosiu, J.~Dafflon, V.~Fernandez, P.~Sanchez, J.~Wolleb, P.~F. Da~Costa, A.~Patel \emph{et~al.}, ``Generative ai for medical imaging: extending the monai framework,'' \emph{arXiv preprint arXiv:2307.15208}, 2023.

\bibitem{song2024exploring}
Y.~Lv, H.~Pan, Z.~Wang, J.~Liang, Y.~Liu, R.~Fu, M.~Liu, Z.~Wang, and B.~Qin, ``Coggpt: Unleashing the power of cognitive dynamics on large language models,'' \emph{arXiv preprint arXiv:2401.08438}, 2024.

\bibitem{huang2024video}
M.~Vukadinovic, X.~Tang, N.~Yuan, P.~Cheng, D.~Li, S.~Cheng, B.~He, and D.~Ouyang, ``Echoprime: A multi-video view-informed vision-language model for comprehensive echocardiography interpretation,'' \emph{arXiv preprint arXiv:2410.09704}, 2024.

\bibitem{wang2024large}
\BIBentryALTinterwordspacing
H.~Yang, Y.~Zhao, Y.~Wu, S.~Wang, T.~Zheng, H.~Zhang, W.~Che, and B.~Qin, ``Large language models meet text-centric multimodal sentiment analysis: {A} survey,'' \emph{CoRR}, vol. abs/2406.08068, 2024. [Online]. Available: \url{https://doi.org/10.48550/arXiv.2406.08068}
\BIBentrySTDinterwordspacing

\bibitem{zhou2023style}
Y.~Zhou and G.~Long, ``Style-aware contrastive learning for multi-style image captioning,'' in \emph{Findings of the Association for Computational Linguistics: EACL 2023}, 2023, pp. 2257--2267.

\bibitem{zhou2021improving}
Y.~Zhou, X.~Geng, T.~Shen, W.~Zhang, and D.~Jiang, ``Improving zero-shot cross-lingual transfer for multilingual question answering over knowledge graph,'' in \emph{Proceedings of the 2021 Conference of the North American Chapter of the Association for Computational Linguistics: Human Language Technologies}, 2021, pp. 5822--5834.

\bibitem{zhou2021modeling}
Y.~Zhou, X.~Geng, T.~Shen, J.~Pei, W.~Zhang, and D.~Jiang, ``Modeling event-pair relations in external knowledge graphs for script reasoning,'' \emph{Findings of the Association for Computational Linguistics: ACL-IJCNLP 2021}, 2021.

\bibitem{zhou2023improving}
Y.~Zhou and G.~Long, ``Improving cross-modal alignment for text-guided image inpainting,'' in \emph{Proceedings of the 17th Conference of the European Chapter of the Association for Computational Linguistics}, 2023, pp. 3445--3456.

\bibitem{zhou2024visual}
Y.~Zhou, X.~Li, Q.~Wang, and J.~Shen, ``Visual in-context learning for large vision-language models,'' in \emph{Findings of the Association for Computational Linguistics, {ACL} 2024, Bangkok, Thailand and virtual meeting, August 11-16, 2024}.\hskip 1em plus 0.5em minus 0.4em\relax Association for Computational Linguistics, 2024, pp. 15\,890--15\,902.

\bibitem{zhou2025weak}
\BIBentryALTinterwordspacing
Y.~Zhou, J.~Shen, and Y.~Cheng, ``Weak to strong generalization for large language models with multi-capabilities,'' in \emph{The Thirteenth International Conference on Learning Representations}, 2025. [Online]. Available: \url{https://openreview.net/forum?id=N1vYivuSKq}
\BIBentrySTDinterwordspacing

\bibitem{kang2024wolf}
\BIBentryALTinterwordspacing
S.~Kang, D.~Kim, J.~Kim, H.~K. Lee, and S.~J. Hwang, ``Wolf: Wide-scope large language model framework for {CXR} understanding,'' \emph{CoRR}, vol. abs/2403.15456, 2024. [Online]. Available: \url{https://doi.org/10.48550/arXiv.2403.15456}
\BIBentrySTDinterwordspacing

\bibitem{lee2023llmcxr}
\BIBentryALTinterwordspacing
S.~Lee, W.~J. Kim, J.~Chang, and J.~C. Ye, ``{LLM-CXR:} instruction-finetuned {LLM} for {CXR} image understanding and generation,'' in \emph{The Twelfth International Conference on Learning Representations, {ICLR} 2024, Vienna, Austria, May 7-11, 2024}.\hskip 1em plus 0.5em minus 0.4em\relax OpenReview.net, 2024. [Online]. Available: \url{https://openreview.net/forum?id=BqHaLnans2}
\BIBentrySTDinterwordspacing

\bibitem{huang2024diffcxr}
\BIBentryALTinterwordspacing
P.~Huang, B.~Guo, S.~Liang, J.~Fu, Y.~Wang, and Y.~Guo, ``Diff-cxr: Report-to-cxr generation through a disease-knowledge enhanced diffusion model,'' \emph{CoRR}, vol. abs/2410.20165, 2024. [Online]. Available: \url{https://doi.org/10.48550/arXiv.2410.20165}
\BIBentrySTDinterwordspacing

\bibitem{han2024advancing}
\BIBentryALTinterwordspacing
W.~Han, C.~Kim, D.~Ju, Y.~Shim, and S.~J. Hwang, ``Advancing text-driven chest x-ray generation with policy-based reinforcement learning,'' in \emph{Medical Image Computing and Computer Assisted Intervention - {MICCAI} 2024 - 27th International Conference, Marrakesh, Morocco, October 6-10, 2024, Proceedings, Part {III}}, ser. Lecture Notes in Computer Science, M.~G. Linguraru, Q.~Dou, A.~Feragen, S.~Giannarou, B.~Glocker, K.~Lekadir, and J.~A. Schnabel, Eds., vol. 15003.\hskip 1em plus 0.5em minus 0.4em\relax Springer, 2024, pp. 56--66. [Online]. Available: \url{https://doi.org/10.1007/978-3-031-72384-1\_6}
\BIBentrySTDinterwordspacing

\bibitem{yao2024addressing}
W.~Yao, C.~Liu, K.~Yin, W.~K. Cheung, and J.~Qin, ``Addressing asynchronicity in clinical multimodal fusion via individualized chest x-ray generation,'' in \emph{Advances in Neural Information Processing Systems 38: Annual Conference on Neural Information Processing Systems 2024, NeurIPS 2024, Vancouver, BC, Canada, December 10 - 15, 2024}, A.~Globersons, L.~Mackey, D.~Belgrave, A.~Fan, U.~Paquet, J.~M. Tomczak, and C.~Zhang, Eds., 2024.

\bibitem{aslani2022optimising}
\BIBentryALTinterwordspacing
S.~Aslani, W.~Lilaonitkul, V.~Gnanananthan, D.~Raj, B.~Rangelov, A.~L. Young, Y.~Hu, P.~Taylor, D.~C. Alexander, and J.~Jacob, ``Optimising chest x-rays for image analysis by identifying and removing confounding factors,'' in \emph{Proceedings of 2022 International Conference on Medical Imaging and Computer-Aided Diagnosis, {MICAD} 2022, Leicester, UK, November 20-21, 2022}, ser. Lecture Notes in Electrical Engineering, R.~Su, Y.~Zhang, H.~Liu, and A.~F. Frangi, Eds., vol. 810.\hskip 1em plus 0.5em minus 0.4em\relax Springer, 2022, pp. 245--254. [Online]. Available: \url{https://doi.org/10.1007/978-981-16-6775-6\_20}
\BIBentrySTDinterwordspacing

\bibitem{castro2024padchestgr}
\BIBentryALTinterwordspacing
D.~C. Castro, A.~Bustos, S.~Bannur, S.~L. Hyland, K.~Bouzid, M.~T. Wetscherek, M.~S{\'{a}}nchez{-}Valverde, L.~Jaques{-}P{\'{e}}rez, L.~P{\'{e}}rez{-}Rodr{\'{\i}}guez, K.~Takeda, J.~M. Salinas, J.~Alvarez{-}Valle, J.~Galant, and A.~Pertusa, ``Padchest-gr: {A} bilingual chest x-ray dataset for grounded radiology report generation,'' \emph{CoRR}, vol. abs/2411.05085, 2024. [Online]. Available: \url{https://doi.org/10.48550/arXiv.2411.05085}
\BIBentrySTDinterwordspacing

\bibitem{han2021ganbased}
\BIBentryALTinterwordspacing
L.~Han, Y.~Lyu, C.~Peng, and S.~K. Zhou, ``Gan-based disentanglement learning for chest x-ray rib suppression,'' \emph{Medical Image Anal.}, vol.~77, p. 102369, 2022. [Online]. Available: \url{https://doi.org/10.1016/j.media.2022.102369}
\BIBentrySTDinterwordspacing

\bibitem{chen2023finematching}
\BIBentryALTinterwordspacing
W.~Chen, L.~Shen, J.~Lin, J.~Luo, X.~Li, and Y.~Yuan, ``Fine-grained image-text alignment in medical imaging enables explainable cyclic image-report generation,'' in \emph{Proceedings of the 62nd Annual Meeting of the Association for Computational Linguistics (Volume 1: Long Papers), {ACL} 2024, Bangkok, Thailand, August 11-16, 2024}, L.~Ku, A.~Martins, and V.~Srikumar, Eds.\hskip 1em plus 0.5em minus 0.4em\relax Association for Computational Linguistics, 2024, pp. 9494--9509. [Online]. Available: \url{https://doi.org/10.18653/v1/2024.acl-long.514}
\BIBentrySTDinterwordspacing

\end{thebibliography}
\end{document}